\documentclass{iaus}
\usepackage{graphicx}

\title[Evolution of twist and dip-shear ] 
{Evolution of twist-shear and dip-shear in flaring active region NOAA 10930}

\author[Gosain, S.  \& Venkatakrishnan, P.]   
{Sanjay Gosain
 \and P. Venkatakrishnan}

\affiliation{Udaipur Solar Observatory, Physical Research Laboratory, \\ P. Box No. 198,
Udaipur 313001, Rajasthan, India \\ email: {\tt sgosain@prl.res.in} \\}

\pubyear{2010}
\volume{xxx}  
\pagerange{119--126}
\jname{IAU Symposium 273: Physics of the Sun and Star Spots}
\editors{Debi Prasad Choudhary \& Klaus G. Strassmeier, eds.}
\begin{document}

\maketitle

\begin{abstract}
We study the evolution of magnetic shear angle in a flare
productive active region NOAA 10930. The magnetic shear angle
is defined as the deviation in the orientation of the observed
magnetic field vector with respect to the potential field
vector. The shear angle is measured in horizontal as well as
vertical plane. The former is computed by taking the difference
between the azimuth angles of the observed and potential field
and is called the twist-shear, while the latter is computed by
taking the difference between the inclination angles of the
observed and potential field and is called the dip-shear. The
evolution of the two shear angles is then tracked over a small
region located over the sheared penumbra of the delta sunspot
in NOAA 10930. We find that, while the twist-shear shows an
increasing trend after the flare the dip-shear shows a
significant drop after the flare. 
\end{abstract}

\section{Introduction}
The non-potential magnetic field in solar active regions stores
the free-energy which is needed to fuel the energetic events
like solar flares. The conventional measure of non-potentiality
has been the so-called magnetic shear angle (Hagyard {\it et
al.} 1984). This angle measures the difference between the
observed and potential field azimuths and has been studied in
relation to the flares (Venkatakrishnan {\it et al.} 1988).
However, this angle measures only the deviations of the
observed field from potential field vector in the horizontal
plane alone. Such deviations are also possible in the vertical
plane i.e., in the magnetic field inclination angles of the
observed and potential field. In order to distinguish between
these two types of shear we call the shear in the horizontal
plane as the twist-shear while the shear in the vertical plane
as the dip-shear.

In this paper we show how the observed magnetic field deviates
from the potential field in the vertical plane in flare
productive active region NOAA 10930. Further, we show the
evolution of these two shear parameters in the penumbral region
located close to the polarity inversion line (PIL) of the delta
sunspot before and after the flare. The high-resolution vector
magnetograms were derived by using the spectropolarimetric
observations from {\it Hinode} Solar Optical Telescope (SOT)
(Tsuneta {\it et al.} 2008). We describe these results in the
following sections.

\section{Observations and Analysis Methods}
The delta sunspot in NOAA 10930 was observed during 12-13
December 2006 by {\it Hinode} space mission (Kosugi {\it et al.
2007)}. The spectropolarimetric data was obtained from {\it
Hinode} SOT/SP instrument (Ichimoto {\it et al.} 2008) and was
reduced and calibrated using SolarSoft package. The calibrated
spectropolarimetric data was then inverted using MERLIN
inversion code (Lites {\it et al.} 2007) at HAO, Boulder, USA.
This inversion code performs the non-linear least squares
fitting of the observed Stokes profiles with the theoretical
Stokes profiles computed under Milne-Eddington model atmosphere
assumptions. The resulting best-fit magnetic parameters were
then resolved for 180 degree azimuth ambiguity by using the
acute angle method. These were then transformed into
heliographic coordinates using the method of Venkatakrishnan
{\it et al.} (1988). The potential field was computed using the
method of Alissandrakis (1981). The magnetograms were
registered by applying the image cross-correlation method on
the continuum intensity images. The figure 1 shows the
continuum intensity maps of the six magnetograms obtained
during 12-13 December 2006. The black rectangular box is the
location where we monitor the dip-shear and the twist-shear.
The figure 2 shows the evolution of the shear parameters inside
this box during the observations.
\section{Results and Discussions}
It can be clearly noticed that: (i) the twist-shear and
dip-shear are correlated i.e., the pixels with large twist
shear also tend to have large value of dip-shear and
vice-versa, with some spread in either parameter, (ii) the
dip-shear shows an increasing trend before the flare, (iii) the
dip-shear decreases significantly after the flare, (iv) the
twist-shear increases after the flare which was also observed
by Jing {\it et al.} (2008).

Any flare related change in the observed  parameters of the
active regions is useful in order to understand the nature of
the energy build-up and its subsequent release in flares and
CMEs. The changes in the line-of-sight magnetic field was been
studied by Sudol \& Harvey (2005) in large number of powerful
flares and firmly established that there is abrupt and
permanent flare related change in active regions. The present
study tries to establish those results on more firm footing by
detecting
 the changes in the magnetic field vector directly. However,
 the slow cadence of the {\it Hinode} SOT/SP observations
 present the biggest limitation in moving forward with such
 studies. We plan to conduct similar study in near future by using
 high-cadence vector magnetograms from the recently launched
 Helioseismic and Magnetic Imager (HMI) onboard Solar Dynamics
 Observatory (SDO).

\begin{figure}    
\centerline{\includegraphics[width=0.85\textwidth,clip=]{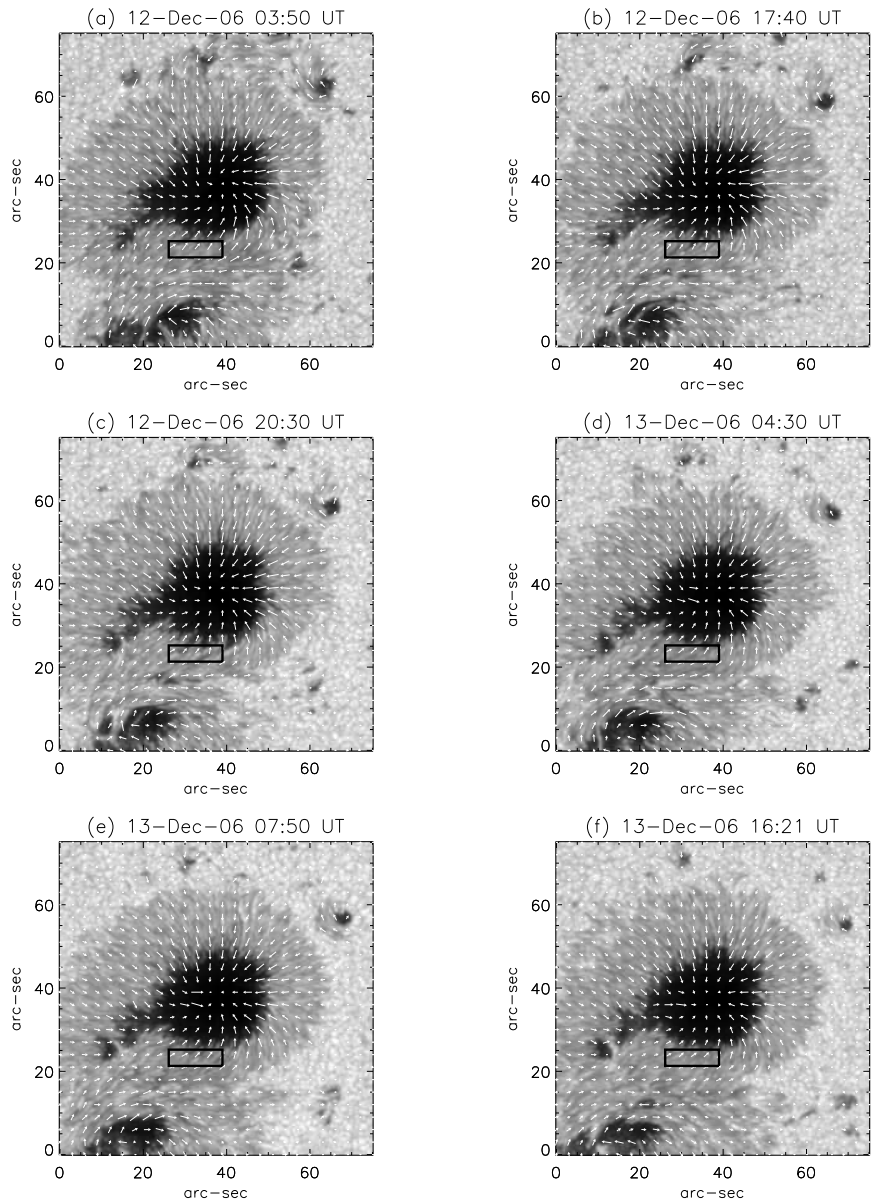}}
\caption{Panels (from top to bottom) show a continuum intensity map of the delta-sunspot in NOAA 10930 during the times mentioned at the top.
The transverse
magnetic field vectors are shown by arrows overlaid upon these maps. The black rectangle, as shown in all panels, is the region where we monitor the evolution of the
twist-shear and dip-shear.}
\end{figure}

\begin{figure}    
\centerline{\includegraphics[width=.85\textwidth,clip=]{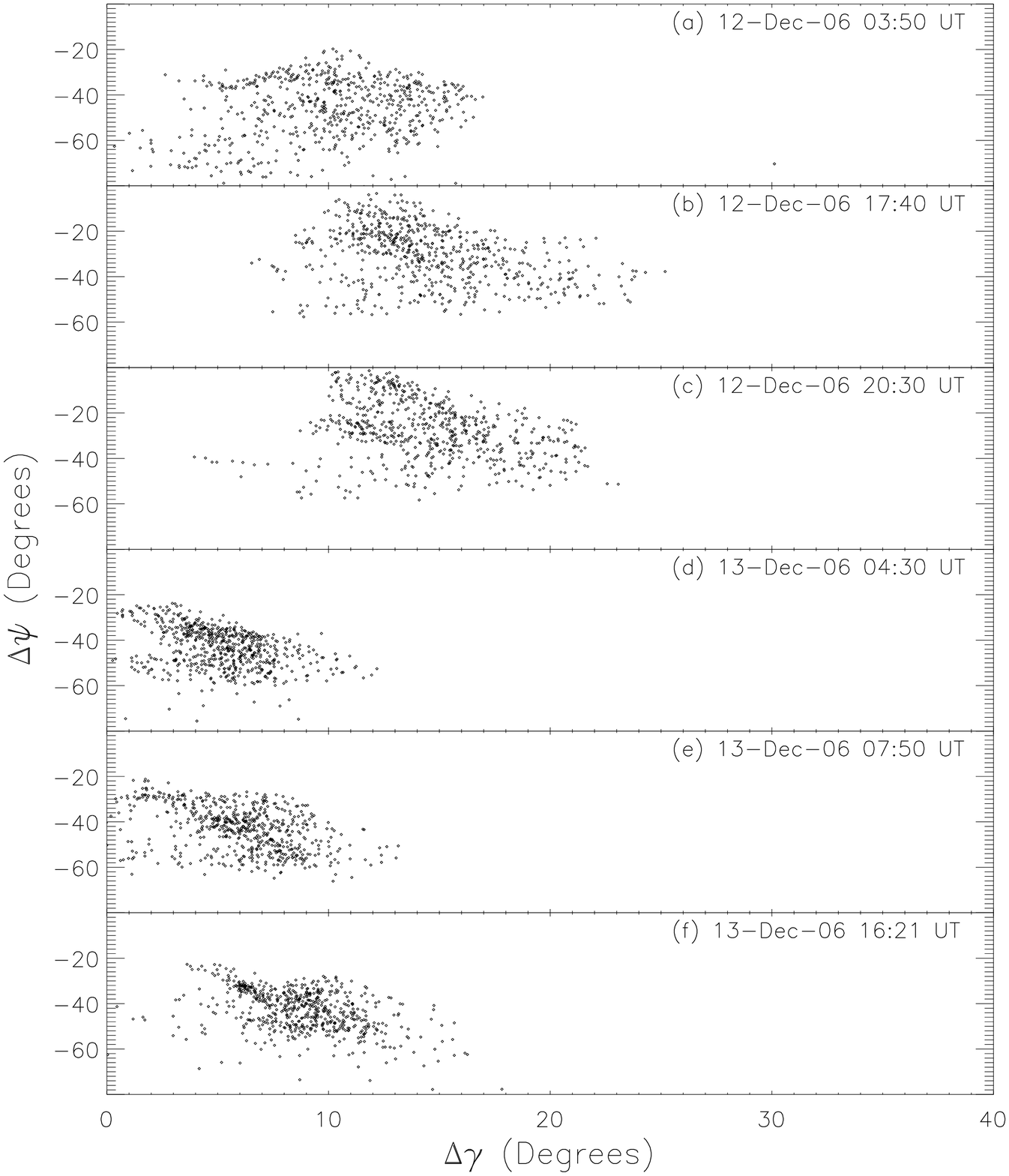}}
\caption{Panels (from top to bottom) show the evolution of the twist-shear and the dip-shear inside the region marked by the rectangle in figure 1. }
\end{figure}

\section{Acknowledgements}
The presentation of this paper in the IAU Symposium 273 was
possible due to financial support from the National Science
Foundation grant numbers ATM 0548260, AST 0968672 and NASA -
Living With a Star grant number 09-LWSTRT09-0039. Hinode is a
Japanese mission developed and launched by ISAS/JAXA, with NAOJ
as domestic partner and NASA and STFC (UK) as international
partners. It is operated by these agencies in co-operation with
ESA and NSC (Norway).

\end{document}